\documentclass[sigconf]{acmart}

\setcopyright{ccbysawithspacing} 

\usepackage{listings}
\usepackage{xcolor}
\usepackage{microtype}
\usepackage{amsmath}
\usepackage{amsthm}
\usepackage{amssymb}
\usepackage{amsfonts}
\usepackage{graphicx}
\usepackage{mathpartir}
\usepackage{mathpartir_extra}
\usepackage{xspace}
\usepackage{array}
\usepackage{proof}
\usepackage{letltxmacro}
\usepackage{enumitem}
\usepackage{todonotes}
\usepackage{notations}

\usepackage{hyperref}
\hypersetup{colorlinks=true,citecolor=green!60!black,urlcolor=blue}

\newcommand{\lts}{$\lambda$-tree syntax\xspace}
\newcommand{\ulc}{untyped $\lambda$-calculus\xspace}
\newcommand{\lP}{$\lambda$Prolog\xspace}
\newcommand{\Ll}{$L_\lambda$\xspace}

\newcommand{\ra}{\rightarrow}

\newcommand{\lam}{\hbox{lam}}
\newcommand{\Let}[2]{\hbox{let}~#1~#2}

\newcommand{\match}[2]{\hbox{match}~#1~#2}
\newcommand{\clause}[3]{\hbox{clause}~#1~#2~#3}

\newcommand{\eval}[2]{#1\mathrel\Downarrow#2}

\newcommand{\matches}[2]{\hbox{matches}~#1~#2}

\newcommand{\back}[1]{\hbox{backslash}~#1}
\newcommand{\arob}[2]{\hbox{arobase}~#1~#2}

\newcommand{\fix}[1]{\hbox{fix}~#1}

\newcommand{\new}[1]{\hbox{new}~#1}

\newcommand{\cons}[2]{#1 \hbox{::} #2}

\newcommand{\all}{\hbox{all}~}
\newcommand{\nab}{\hbox{nab}~}
\newcommand{\Gee}{{\mathcal G}}

\newcommand{\mlts}{\hbox{\textsc{MLTS}}\xspace}
\newcommand{\bfz}{{\bf 0}\xspace}

\newcommand{\judty}[3]{#1\vdash \lsti{#2} : \lsti{#3}}
\newcommand{\judr }[4]{#1\vdash \lsti{#2} : \lsti{#3} : \lsti{#4}}
\newcommand{\judlh}[4]{#1\vdash \lsti{#2} : \lsti{#3} \dashv #4}
\newcommand{\openty}[1]{\hbox{open~}\lsti{#1}}

\colorlet{lprolog}{blue!70!black}
\colorlet{abellatop}{blue!70!green}
\colorlet{abellatac}{orange!30!black}
\colorlet{abellabad}{red!80!yellow}
\definecolor{bleu}{HTML}{000DB3}

\lstdefinelanguage{lprolog}{%
  alsoletter={-},
  classoffset=0,%
  morekeywords={sig,module,type,kind,pi,sigma,end},%
  keywordstyle=\color{lprolog},%
  classoffset=0,%
  otherkeywords={:-,\\,=,=>,<=,\&},%
  sensitive=true,%
  morestring=[bd]",%
  morecomment=[l]\%,%
  morecomment=[n]{/*}{*/},%
}

\lstdefinelanguage{abella}[]{lprolog}{%
  alsoletter={-},
  classoffset=1,%
  morekeywords={Close,CoDefine,Define,Kind,Query,Quit,Specification,
    Set,Split,Theorem,Type,Undo,by,as,prop,true,false,forall,exists,nabla},%
  keywordstyle=\color{abellatop},%
  classoffset=2,%
  morekeywords={abbrev,apply,backchain,case,coinduction,cut,
    induction,inst,intros,monotone,on,permute,rename,left,right,witness,
    search,split,to,unabbrev,unfold,assert,with},%
  keywordstyle=\color{abellatac},%
  classoffset=3,%
  morekeywords={undo,abort,skip,clear},%
  keywordstyle=\color{abellabad}\underbar,%
  classoffset=0,%
}

\lstdefinelanguage{mlts}[Objective]{Caml}{%
  morekeywords={nab, new},%
  otherkeywords={@,\\, =>, =>},%
}

\newcommand*{\SavedLstInline}{}
\LetLtxMacro\SavedLstInline\lstinline
\DeclareRobustCommand*{\lstinline}{%
  \ifmmode
    \let\SavedBGroup\bgroup
    \def\bgroup{%
      \let\bgroup\SavedBGroup
      \hbox\bgroup
    }%
  \fi
  \SavedLstInline
}
\DeclareRobustCommand\lsti[1][]{%
  \lstinline[language=mlts,basicstyle=\ttfamily,keepspaces=true,#1]}
\DeclareRobustCommand\lstip[1][]{%
  \lstinline[language=lprolog,basicstyle=\ttfamily,keepspaces=true,#1]}

\renewcommand{\cite}[1]{\citep*{#1}}

\theoremstyle{plain}

\title{Functional programming with \lts}
\author{Ulysse G\'erard, Dale Miller, and Gabriel Scherer}
\affiliation{%
  \institution{Inria \& LIX, Ecole Polytechnique}
  \city{Palaiseau}
  \country{France}}

\begin{document}
\acmConference[PPDP'2019]{21th International Symposium on Principles
  and Practice of Declarative Programming}{October 7--9}{Porto,
  Portugal}

\begin{abstract}
We present the design of a new functional programming language, \mlts,
that uses the \emph{$\lambda$-tree} syntax approach to encoding
bindings appearing within data structures.
In this approach, bindings never become free nor escape their scope:
instead, binders in data structures are permitted to \emph{move} to
binders within programs.
The design of \mlts includes additional sites within programs that
directly support this movement of bindings.
In order to formally define the language's operational semantics, we
present an abstract syntax for \mlts and a natural semantics for its
evaluation.
We shall view such natural semantics as a logical theory within a rich
logic that includes both \emph{nominal abstraction} and the
$\nabla$\emph{-quantifier}: as a result, the natural semantics
specification of \mlts can be given a succinct and elegant
presentation.
We present a typing discipline that naturally extends the typing of
core ML programs and we illustrate the features of \mlts by presenting
several examples.
An on-line interpreter for \mlts is briefly described.
\end{abstract}

\copyrightyear{2019} 
\acmYear{2019} 
\acmConference[PPDP '19]{Principles and Practice of Programming Languages 2019}{October 7--9, 2019}{Porto, Portugal}
\acmBooktitle{Principles and Practice of Programming Languages 2019 (PPDP '19), October 7--9, 2019, Porto, Portugal}
\acmPrice{}
\acmDOI{10.1145/3354166.3354177}
\acmISBN{978-1-4503-7249-7/19/10}

\ccsdesc[300]{Software and its engineering~Functional languages}
\ccsdesc[300]{Software and its engineering~Data types and structures}


\maketitle

\section{Introduction}\label{sec:intro}

Even from the earliest days of high-level programming, functional
programming languages were used to build systems that manipulate the
syntax of various programming languages and logics.
For example, Lisp was a common language for building theorem provers,
interpreters, compilers, and parsers, and the ML programming language
was designed as a ``meta-language'' for a proof checker~\cite{gordon79}.
While these various tasks involve the manipulation of syntax, none of
these earliest functional programming languages provided support for a
key feature of almost all programming languages and logics: variable
binding.

Bindings in syntactic expressions have been given, of course, a range
of different treatments within the functional programming setting.
Common approaches are to implement bindings by using variable names
or, in a more abstract way, by using de Bruijn
indexes~\cite{debruijn72}.
Since such techniques are quite complex to get right and since
bindings are so pervasive, a great deal of energy has gone into making
tools and libraries that can help deal with binders: for example,
there is the \emph{locally nameless} approach
\cite{chargueraud11jar,gordon94tphol,mcbride04haskell} and the
\emph{parametric higher-order abstract syntax}
approach~\cite{chlipala08icfp}.

Extending a functional programming language with features that support
bindings in data has been considered before: for example, there have
been the FreshML~\cite{gabbay03icfp,pottier07lics} and C$\alpha$ML
\cite{pottier06ml} extensions to ML-style functional programming
languages.
Also, entirely new functional programming languages, such as the
dependently typed Beluga~\cite{pientka10ijcar} language, have been
designed and implemented with the goal of supporting bindings in syntax.
In the domains of logic programming and theorem proving, several
designs and implemented systems exist that incorporate approaches to
binding: such systems include Isabelle's generic reasoning
core~\cite{paulson89jar}, \lP~\cite{nadathur88iclp,miller12proghol},
Qu-Prolog~\cite{cheng91iclp}, Twelf~\cite{pfenning99cade},
$\alpha$Prolog~\cite{cheney04iclp}, the Minlog
prover~\cite{schwichtenberg06provers}, and the Abella theorem
prover~\cite{baelde14jfr}.

In this paper we present \mlts, a new language that extends (the core
of) ML and incorporates the \emph{\lts} approach to
encoding the abstract syntax of data structures containing binders.
Briefly, we can define the \lts approach to syntax as following the
three tenets:
(1)~Syntax is encoded as simply typed $\lambda$-terms in which the
primitive types are identified with syntactic categories.
(2)~Equality of syntax must include $\alpha\beta\eta$-conversion.
(3)~Bound variables never become free: instead, their binding scope
can move.
This latter tenet introduces the most characteristic aspect of \lts
which is often called \emph{binder mobility}.
\mlts is, in fact, an acronym for \emph{mobility and \lts}.

This paper contains the following contributions.
\begin{itemize}[leftmargin=*,label=-]
\item  We present the design of \mlts, a new functional language
  prototype for dealing with bindings, aiming at expressivity and
  generality.

\item We show how the treatment of bindings that has been successful
  in the logic programming and theorem proving systems \lP, Twelf, and
  Abella, can be incorporated into a functional programming language.

  At the same time, \mlts{} remains a ML-family language; nominals are
  treated similarly to constructors of algebraic datatypes (in
  expressions and patterns), distinguishing our design from existing
  proposals, such as Delphin and Beluga.

\item We present some of the metatheory of \mlts.

\item We have a full prototype implementation that is accessible online.
\end{itemize}

This paper is organized as follows.
Section~\ref{sec:mlts} introduces the language \mlts and aims to give
a working understanding to the reader of its new constructs and
current implementation.
Section~\ref{sec:design} presents some of the foundational
aspects of \mlts' design, which comes from the proof-search (logic
programming) paradigm, along with its natural semantics.
Section~\ref{sec:formal} contains a formal description of the typing
system for \mlts as well as some static restrictions we impose
on the language to obtain good reasoning principles.
We also state here some meta-theorems about \mlts.
In Section~\ref{sec:mobility} we elaborate on several
issues that surround the insertion of binder mobility into this
functional programming language.
Finally, in Sections~\ref{sec:future}, ~\ref{sec:related}, and
~\ref{sec:conclusion} we present future work, related work, and
conclude.

\section{A tour of \mlts}\label{sec:mlts}

We chose the concrete syntax of \mlts to be an extension of that of the OCaml
programming language (a program in \mlts not using the new language
features should be accepted by the \lsti|ocamlc| compiler).
We assume that the reader is familiar with basic syntactic conventions
of OCaml~\cite{ocaml.website}, many of which are shared with most ML-like
programming languages.

This section presents the new constructs of \mlts along with a set of
examples. We also provide a web application, TryMLTS~\cite{trymlts}, that can
serve as a companion during the reading of this introduction to the language.

\subsection{The binding features of \mlts}

\textsc{MLTS} contains the following five new language features.
\begin{enumerate}
\item Datatypes can be extended to contain new \emph{nominal}
  constants and the \lsti{(new X in M)} program phrase introduces a fresh
  nominal \lsti{X} in the scope of the evaluation of the term \lsti{M}. The value of this
  expression is the value of \lsti{M}, provided that this value does not contain any
  remaining occurrence of \lsti{X} -- this would be a nominal escape failure. For
  example, the term \lsti{(new X in X)} fails during evaluation.

\item A new typing constructor \lsti{=>} is used to type bindings
  within term structures.  This constructor is an addition to the
  already familiar constructor \lsti{->} used for function types.

\item The \emph{backslash} (\lsti{\} as an infix symbol that
  associates to the right) is used to form an abstraction of a nominal
  over its scope.  For example, \lsti{(X\body)} is a syntactic
  expression that hides the nominal \lsti{X} in the scope \lsti{body}.
  Thus the backslash \emph{introduces} an abstraction.

\item
  The infix symbol \lsti{@} \emph{eliminates} an abstraction: for
  example, the expression \lsti{((X\body) @ t)} denotes the result of
  the capture-avoiding substitution of the abstracted nominal \lsti{X}
  by the term \lsti{t} in \lsti{body}. The notation \lsti{(t @ u v)}
  stands for \lsti{(t @ u @ v)} (\lsti{@} associates to the left).

\item Clauses within match-expressions can also contain the
  \lsti{(nab X in p -> t)}
  binding form. Here, \lsti{X} is a nominal local to the clause \lsti|p -> t|.
  At runtime it will be substituted by a nominal \lsti{Y} from the ambient
  context that appears in the scrutiny of the match at the same position than
  \lsti{X} does in \lsti{p} but does not already appear in \lsti|p -> t|.
  %
\end{enumerate}
These new term operators have the following precedence from highest to lowest:
\lsti|@|, \lsti|new| and \lsti|\|. Other operators have the same precedences
and associativity than in OCaml. Thus the expression
\lsti|fun r -> X\ new Y in r @ X| reads as:
\lsti|fun r -> (X\ (new Y in (r @ X)))|.
All three binding expressions---\lsti{(X\body)},
\lsti{(new X in body)} and \lsti{(nab X in rule)}---are subject to
$\alpha$-renaming of bound variables, just as the names of variables
bound in \lsti{let} declarations and function definitions.
As we shall see, nominals are best thought of as constructors: as a
consequence, we follow the OCaml convention of capitalizing their names.
We are assuming that, in all parts of \mlts, the names of nominals (of
bound variables in general) are not available to programs since
$\alpha$-conversion (the alphabetic change of bound variables) is
always applicable.
Thus, compilers are free to implement nominals in any number of ways,
even ways in which they do not have, say, print names.

We enforce a few restrictions (discussed in
Section~\ref{ssec:unification}) on match expressions: Every
\lsti{nab}-bounded nominals must occur rigidly (defined
in Section~\ref{subsubsec:rigid-occurrence}) in the pattern  and
expressions of the form \lsti{(m @ X1 ... Xj)} in patterns are
restricted so that \lsti{m} is a pattern variable and \lsti{X1},
$\ldots$, \lsti{Xj} are distinct nominals bound within the scope of
the pattern binding on \lsti{m} (which, as a pattern variable, is
scoped outside the scopes of \lsti{nab}-bound nominals and over the
whole rule).
This restriction is essentially the same as the one required by
\emph{higher-order pattern unification}~\cite{miller91jlc}: as a
result, pattern matching in this setting is a simple generalization of
usual first-order pattern matching.

We note that the expression \lsti{(X\ r @ X)} is interchangeable with
the simple expression \lsti{r}: that is, when \lsti{r} is of \lsti{=>}
type, $\eta$-equality holds.

We now present two series of examples of \mlts programs.
We hope that the informal presentation given above plus the
simplicity of the examples will give a working understanding of the
semantics of \mlts.
We delay the formal definition of the operational semantics of \mlts
until Section~\ref{sec:natsem}.

\subsection{Examples: the \ulc}\label{sec:examples}

The untyped $\lambda$-terms can be defined in \mlts as the datatype:
\begin{lstlisting}[linerange={1-3}]
type tm =
     | App of tm * tm
     | Abs of tm => tm ;;
\end{lstlisting}
The use of the \lsti{=>} type constructor here indicates that the
argument of \lsti{Abs} is a \emph{binding abstraction} of a \lsti{tm} over a
\lsti{tm}. Notice the absence of clause for variables.
In \mlts, such a type, called an \emph{open} type, can be extended with a
collection of nominal constructors of type \lsti{tm}.
Just as the type \lsti{tm} denotes a syntactic category of untyped
$\lambda$-terms, the type  \lsti{tm => tm} denotes the syntactic
category of terms abstracted over such terms.

Following usual conventions, expressions whose concrete syntax have
nested binders using the same name are disambiguated by the parser by
linking the named variable with the closest binder.
Thus, the concrete syntax \lsti{(Abs(X\ Abs(X\ X)))} is parsed as a term
$\alpha$-equivalent to \lsti{(Abs(Y\ Abs(X\ X)))}.
Similarly, the expression \lsti{(let n = 2 in let n = 3 in n)} is
parsed as an expression $\alpha$-equivalent to
\lsti{(let m = 2 in let n = 3 in n)}: this expression has value \lsti{3}.

\begin{figure}
\begin{lstlisting}[linerange={5-9}]
let rec size t =
    match t with
	| App(n, m)  -> 1 + size n + size m
	| Abs(r) -> 1 + new X in size (r @ X)
	| nab X in X -> 1;;
\end{lstlisting}
\caption{A program for computing the size of a $\lambda$-term.}
\label{fig:size}

\begin{lstlisting}[linerange={5-28}]
let subst t u = new X in
    let rec aux t = match t with
      | X -> u
      | nab Y in Y -> Y
      | App(u, v) -> App(aux u, aux v)
      | Abs r -> Abs(Y\ aux (r @ Y))
    in aux (t @ X);;

let rec beta t = match t with
  | nab X in X -> X
  | Abs r      -> Abs(Y\ beta (r @ Y))
  | App(m, n)  ->
    let m = beta m in let n = beta n in
    begin
        match m with
        | Abs r -> beta (subst r n)
        | _ -> App(m, n)
    end ;;

let two  = Abs(F\ Abs(X\ App(F, App(F, X))));;
let plus = Abs(M\ Abs(N\ Abs(F\ Abs(X\
           App(App(M,F), App(App(N,F),X))))));;
let times = Abs(M\ Abs(N\ Abs(F\ Abs(X\ 
              App(App(M, App(N, F)), X)))));;
\end{lstlisting}\noindent
\caption{The function that computes the substitution $[t/x]u$ and the
  (partial) function that computes the $\beta$-normal form of its argument.}
\label{fig:beta}
\end{figure}

The \mlts program in Figure~\ref{fig:size} computes the size of an
untyped $\lambda$-term \lsti|t|.
For example, \lsti{(size (App(Abs(X\X), Abs(X\X))))} evaluates to
5.
In the second match rule, the match-variable \lsti{r} is bound
to an expression built using the backslash.
On the right of that rule, \lsti{r} is applied to a single argument
which is a newly provided nominal constructor of type \lsti{tm}.
The third match rule contains the \lsti{nab} binder that allows the
token \lsti{X} to match any nominal: alternatively, that last clause
could have matched any non-\lsti{App} and non-\lsti{Abs} term by using
the clause \lsti{| _ -> 1}.
(Note that as written, the three match rules used to define
\lsti{size} could have been listed in any order.)
The following sequence of expressions shows the evolution of
a computation involving the \lsti{size} function.
\begin{lstlisting}[linerange={13-18}]
size (Abs (X\ (Abs (Y\ (App(X,Y))))));;
1 + new X in size (Abs (Y\ (App(X,Y))));;
1 + new X in 1 + new Y in size (App(X,Y));;
1 + new X in 1 + new Y in 1 + size X + size Y;;
1 + new X in 1 + new Y in 1 + 1 + 1;;
\end{lstlisting}
The first call to \lsti{size} binds the pattern variable \lsti{r}
to \lsti{X\ Abs(Y\ App(X,Y))}.
It is important to note that the names of bound variables within \mlts
programs and data structures are fictions: in the expressions above,
binding names are chosen for readability.
%


Figure~\ref{fig:beta} defines the function \lsti{(subst t u)} that
takes an abstraction over terms
\lsti{t} and a term \lsti{u} and returns the result of substituting
the (top-level) bound variable of \lsti{t} with \lsti{u}.
This function works by first introducing a new nominal \lsti{X}
and then defining an auxiliary function that replaces that nominal in a
term with the term \lsti{u}.
Finally, that auxiliary function is called on the expression
\lsti{(t @ X)} which is the result of ``moving'' the top-level bound variable
in \lsti{t} to the binding occurrence of the expression \lsti{new X in}.
(As we note in Section~\ref{ssec:cost}, such binder
movement can sometimes be implemented in constant time.)
This substitution function has the type
\lsti{(tm => tm) -> (tm -> tm)}: that is, it is used to inject the
abstraction type \lsti{=>} into the function type \lsti{->}.
Substitution is then used by the second function of Figure~\ref{fig:beta}, \lsti|beta|, to
compute the $\beta$-normal form of a given term of type \lsti{tm}.
This figure also contains the Church numeral for 2 and operations for
addition and multiplication on Church numerals.  In the resulting
evaluation context, the values computed by
\lsti{(beta (App(App(plus, two), two)))} and
\lsti{(beta (App(App(times, two), two)))} are both the Church numeral
for 4.


\begin{figure}[t]
\begin{lstlisting}[linerange={19-37}]
let rec vacp1 t =  match t with
    | X\X                 -> false
    | nab Y in X\ Y       -> true
    | X\ App(m @ X, n @ X)-> vacp1 m && vacp1 n
    | X\ Abs(Y\ r @ X Y)  -> new Y in
                             vacp1 (X\ r @ X Y);;

let rec vacp2 t =
    new X in
    let rec aux term = match term with
        | X          -> false
        | nab Y in Y -> true
        | App(m, n)  -> aux m && aux n
        | Abs(u)     -> new Y in aux (u @ Y)
    in aux (t @ X);;

let vacp3 t = match t with
    | X\s  -> true
    | _ -> false;;
\end{lstlisting}
\caption{Three implementations for determining if an abstraction is vacuous.}
\label{fig:vacp}
\begin{lstlisting}[linerange={5-15}]
let rec assoc x alist = match alist with
  | ((u,y)::alst) -> if (u = x) then y else (assoc x alst);;

type tm' =
     | App' of tm' * tm'
     | Abs' of tm' => tm';;

let rec id gamma term = match term with
  | App(m,n)   -> App'(id gamma m,id gamma n)
  | Abs(r)     -> new X in Abs'(Y\ (id ((X,Y)::gamma) (r @ X)))
  | nab X in X -> assoc X gamma;;
\end{lstlisting}\noindent
\caption{Translating from \lsti{tm} to its mirror version  \lsti{tm'}.}
\label{fig:simple}
\end{figure}

For another example, consider a program that returns \lsti{true}
if and only if its argument, of type \lsti{tm => tm},
is such that its top-level bound variable is a vacuous binding (i.e.,
the bound variable is not free in its scope).
Figure~\ref{fig:vacp} contains three implementations of this
boolean-valued function. The first implementation proceeds
by matching patterns with the prefix \lsti{X\}, thereby, matching
expressions of type \lsti{tm => tm}.
The second implementation uses a different style: it creates a new
nominal \lsti{X} and proceeds to work on the term \lsti{t @ X}, in the
same fashion as the \lsti|size| example.
The internal \lsti{aux} function is then defined to search for occurrences
of \lsti{X} in that term.
The third implementation, \lsti{vacp3}, is not (overtly) recursive since the
entire effort of checking for the vacuous binding is done during
pattern matching.
The first match rule of this third implementation is essentially
asking the question: is there an instantiation for the (pattern)
variable $s$ so that the $\lambda X.s$ equals $t$?
This question can be posed as asking if the logical formula $\exists
s. (\lambda X.s) = t$ can be proved.
In this latter form, it should be clear that since substitution is
intended as a logical operation, the result of substituting for $s$
never allows for variable capture.
Hence, every instance of the existential quantifier yields an
equation with a left-hand side that is a vacuous abstraction.
Of course, this kind of pattern matching requires a recursive analysis
of the term $t$ and that can make pattern matching
costly.
To address that cost, pattern matching can be restricted so
that such patterns do not occur (see Section~\ref{sec:future}) or
static checks can be added that often make such recursive descents
unnecessary (see Section~\ref{ssec:cost}).


For another simple example of computing on the untyped
$\lambda$-calculus, consider introducing a mirror version of
\lsti{tm}, as is done in Figure~\ref{fig:simple}, and
writing the function that constructs the mirror term in \lsti{tm'}
from an input term \lsti{tm}.
This computation is achieved by adding a context (an association list)
as an extra argument that maintains the association of bound variables
of type \lsti{tm} and those of type \lsti{tm'}.
The value of \lsti{id [] (Abs(X\ Abs(Y\ App(X,Y))))} is
\lsti{(Abs'(X\ Abs'(Y\ App'(X,Y))))} (the types of \lsti{X} and
\lsti{Y} in these two expressions are, of course, different).


\begin{figure}[t]
\begin{lstlisting}[linerange={5-33}]
type deb =
     | Dapp of deb * deb
     | Dabs of deb
     | Dvar of int;;

let rec nth n l = match (n, l) with
  | (0, x::k) -> x
  | (c, x::k) -> nth (c - 1) k;;

let index x l = 
    let rec aux c x k = match (x, k) with
          | nab X   in (X, X::(l @ X)) -> c
          | nab X Y in (X, Y::(l @ X Y)) -> 
                      aux (c + 1) x (l @ X Y)
     in aux 0 x l;;

let rec trans prefix term = match term with
  | App(m, n)  -> Dapp(trans prefix m, 
                       trans prefix n)
  | Abs r      -> new X in 
                  Dabs(trans (X::prefix) (r @ X))
  | nab Y in Y -> Dvar (index Y prefix);;

let rec dtrans prefix term = match term with
  | Dapp(m, n) -> App(dtrans prefix m, 
                      dtrans prefix n)
  | Dabs r     -> Abs(X\ dtrans (X::prefix) r)
  | Dvar c     -> nth c prefix;;
\end{lstlisting}
\caption{De Bruijn's style syntax and its conversions with
  type \lsti{tm}.}
\label{fig:deb}
\end{figure}

Figure~\ref{fig:deb} presents a datatype for the untyped
$\lambda$-calculus in De Bruijn's style%
~\cite{debruijn72} as well as the functions that can convert between
that syntax and the one with explicit bindings.
The auxiliary functions \lsti{nth} and \lsti{index} take a list of
nominals as their second argument: \lsti{nth} takes also an integer
\lsti{n} and returns the $n^{th}$ nominal in that list while
\lsti{index} takes a nominal and returns its ordinal position in that
list.
For example, the value of
\begin{lstlisting}[linerange={39-39}]
trans [] (Abs(X\ Abs(Y\ Abs(Z\ App(X,Z)))));;
\end{lstlisting}
is the term
\lsti{Dabs(Dabs(Dabs(Dapp(Dvar 2, Dvar 0))))} of type \lsti{deb}.
If \lsti{dtrans []} is applied to this second term, the former term is
returned (modulo $\alpha$-renaming, of course).



\subsection{Examples: Higher-order programming}\label{sec:hop}

\begin{figure}[t]
\begin{lstlisting}[language=mlts,linerange={11-32}]
let rec maptm fapp fabs fvar term =
  match term with
   | App(m,n) -> fapp (maptm fapp fabs fvar m)
                      (maptm fapp fabs fvar n)
   | Abs(r) -> fabs (fun x -> 
      match x with 
            | nab X in X ->
              maptm fapp fabs fvar (r @ X))
   | nab X in X -> fvar X;;

let mapvar fvar term = 
    maptm (fun m -> fun n -> App(m, n)) 
          (fun r -> Abs(X\ r X))
          fvar term;;

let lookup sub var = match var with 
  | nab X in X -> 
    let rec aux s = match s with 
                    | []         -> X
                    | (X,t)::sub -> t
                    | (y,t)::sub -> aux sub
     in aux sub;;
\end{lstlisting}
\begin{lstlisting}[language=mlts,linerange={53-72}]
     
let rec remove x l = match l with
    | []    -> []
    | h::tl -> if h = x then remove x tl 
                        else h::(remove x tl);;
	      
let fv term = 
    maptm union
         (fun r -> new X in remove X (r X))
         (fun x -> x::[]) term;;

let size term = 
    maptm (fun x -> fun y -> 1 + x + y)
          (fun r -> new X in 1 + (r X)) 
          (fun x -> 1) term;;

let terminals term = 
    maptm (fun x -> fun y -> x + y)
          (fun r -> new X in (r X))
          (fun x -> 1) term;;
\end{lstlisting}
\caption{Various computations on untyped $\lambda$-terms using
  higher-order programs.  Note that there are several occurrences of
  \lsti{(r X)} above that should not be written as \lsti{(r @ X)}.}
\label{fig:lookup}
\end{figure}

Recall the familiar higher-order function ``fold-right''.
\begin{lstlisting}[language=mlts,linerange={1-3}]
let rec foldr f a lst = match lst with
  | [] -> a
  | x :: xs -> f x (foldr f a xs);;
\end{lstlisting}
This function can be viewed as replacing all occurrences of \lsti{::}
with the binary function \lsti{f} and all occurrences of \lsti{[]}
with \lsti{a}.
The higher-order program \lsti{maptm} in Figure~\ref{fig:lookup} does
the analogous operation on the datatype of untyped $\lambda$-terms
\lsti{tm}.
In particular, the constructors \lsti{App} and \lsti{Abs} are replaced
by functions \lsti{fapp} and \lsti{fabs}, respectively.  In addition,
the function \lsti{fvar} is applied to all nominals encountered in the
term.
This higher-order function can be used to define a number of other
useful and familiar functions.
For example, \lsti{mapvar} function is a specialization of the
\lsti{maptm} function that just applies a given function to all
nominals in an untyped $\lambda$-term.
The application of a substitution (an expression of type
\lsti{(tm * tm) list}) to a term of type \lsti{tm} can then be seen as
the result of applying the \lsti{lookup} function to every variable in
the term (using \lsti{mapvar}).
Using the functions in Figure~\ref{fig:lookup},
the three expressions
\begin{lstlisting}[language=mlts,linerange={34-39}]
Abs(X\ mapvar (fun x -> X)
              (Abs(U\ Abs(V\ App(U,V)))));;
new X in new Y in lookup ((X,Abs(U\U))::
              (Y, Abs(U\ App(U,U)))::[]) X;;
new X in new Y in lookup ((X,Abs(U\U))::
               (Y, Abs(U\ App(U,U)))::[]) Y;;
\end{lstlisting}
\noindent evaluate to the following three $\lambda$-terms.
\begin{lstlisting}
Abs(X\ Abs(Y\ Abs(Z\ App(X, X))))
Abs(X\ X)
Abs(X\ App(X, X))
\end{lstlisting}

Three additional functions are defined in Figure~\ref{fig:lookup}:
\lsti{fv} constructs the list of free variables in a term;
\lsti{size} is a re-implementation of the \lsti{size} function
presented in Section~\ref{sec:examples}; and
\lsti{terminals} counts the number of variable occurrences (terminal
nodes) in its argument.


\subsection{Current prototype implementation}
We have a prototype implementation of \mlts.
A parser from our extended OCaml syntax and a transpiler that
generates \lP code are implemented in OCaml.
A simple evaluator and type checker written in \lP are used to
type-check and execute the generated \mlts code.
The implementation of the evaluator in \lP is rather compact but not
completely trivial since the natural semantics of \mlts (presented in
Section~\ref{sec:natsem}) contains features (namely,
$\nabla$-quantification and nominal abstraction) that are not native
to \lP: they needed to be implemented.
Both the Teyjus~\cite{teyjus.website} and the Elpi~\cite{dunchev15lpar}
implementations of \lP can be used to execute the \mlts interpreter.
Since Elpi, the parser, and the
transpiler are written in OCaml, web-based execution was made possible
by compiling the OCaml bytecode to a Javascript client library with
\lsti{js_of_ocaml}~\cite{js-of-ocaml}.

There is little about this prototype implementation that is focused on
providing an efficient implementation of \mlts.
Instead, the prototype is useful for exploring the exact meaning and
possible uses of the new program features.

\section{The logical foundations of a semantic definition of \mlts}
\label{sec:design}

Bindings are such an intimate part of the nature of syntax that we
should expect that our high-level programming languages account for
them directly: for example, any built-in notion of equality or
matching should respect at least $\alpha$-conversion.
(The paper~\cite{miller18jar} contains an extended argument of this
point in the setting of logic programming and proof assistants.)
Another reason to include binders as a primitive within a functional
programming languages is that their semantics have a well understood
declarative and operational treatment.
For example, Church's higher-order logic STT~\cite{church40} contains
an elegant integration of bindings in both terms and formulas.
His logic also identifies equality for both terms and formulas with
$\alpha\beta\eta$-conversion.
Church's integration is also a popular one in theorem proving---being
the core logic of the Isabelle~\cite{paulson94book}, HOL
\cite{harrison09hol,gordon91tphol}, and Abella~\cite{baelde14jfr}
theorem provers---as well as the logic programming language \lP
\cite{miller12proghol}.
Given the existence of these provers, a good literature now exists
that describes how to effectively implement STT and closely related
logics.
Since the formal specifications of evaluation and typing will be given
using inference rules and since such rules can be viewed
as quantified formulas, this literature provides means for
implementing \mlts.

\subsection{Equality modulo $\alpha$, $\beta$, $\eta$ conversion}

The abstract syntax behind \mlts is essentially a simply typed
$\lambda$-term that encodes \ulc, as described in
Section~\ref{sec:natsem}.
Furthermore, the equality theory of such terms is given by the
familiar $\alpha$, $\beta$, $\eta$ conversion rules.
As a result, a programming language that adopts this notion of
equality cannot take an abstraction and return, say, the name of its
bound variable: since that name can be changed via the
$\alpha$-conversion, such an operation would not be a proper
function.
Thus, it is not possible to decompose the untyped $\lambda$-term
$\lambda x.t$ into the two components $x$ and $t$.
Not being able to retrieve a bound variable's name might appear as a
serious deficiency but, in fact, it can be a valuable feature of the
language: for example, a compiler does not need to maintain such names
and can choose any number of different, low-level representations of
bindings to exploit during execution.
Since the names of bindings seldom have semantically meaningful value,
dropping them entirely is an interesting design choice.
That choice is similar to one taken in ML-style languages in which the
location in memory of a reference cell is not maintained as a value in
the language.

The relation of $\lambda$-conversion is invoked when evaluating
the expression \lsti{(t @ s1 ... sn)}.
As we shall see, \mlts is a typed language so we can assume that the
expressions $\lsti{s1},\ldots,\lsti{sn}$ have types
$\gamma_1,\ldots,\gamma_n$, respectively, and that \lsti{t} must have
type $\gamma_1\Rightarrow\cdots\Rightarrow\gamma_n\Rightarrow\gamma_0$.
Thus, \lsti{t} is $\eta$-equivalent to a term with $n$ abstractions,
for example, \lsti{X1\...Xn\ t'} and the value of the expression
\lsti{(t @ s1 ... sn)} is the result of performing
$\lambda$-normalization of \lsti{((X1\...Xn\ t') s1 ... sn)}.

\subsection{Match rule quantification}

Match rules in \mlts contain two kinds of quantification.
The familiar quantification of pattern variables can
be interpreted as being universal quantifiers.
For example, the first rule defining the \lsti{size} function in
Section~\ref{sec:examples}, namely,
\begin{lstlisting}
	| App(n, m)  -> 1 + size n + size m
\end{lstlisting}
can be encoded as the logical statement
\[
\forall \lsti{m}\forall \lsti{n} [
  \lsti{(size (App(n, m)))} =
  \lsti{ 1 + size n + size m}].
\]
The third match rule for \lsti{size} contains the
binder \lsti{nab}
\begin{lstlisting}
	| nab X in X -> 1
\end{lstlisting}
which corresponds approximately to the \emph{generic}
$\nabla$-quantifier (pronounced nabla) that is found in various
efforts to formalize the metatheory of computational
systems (see~\cite{miller05tocl,baelde14jfr} and
Section~\ref{sec:natsem}).
That is, this rule can be encoded as $\nabla
\lsti{x}. (\lsti{size x} = \lsti{1})$: that is, the size of a nominal
constant is 1.

Although there are two kinds of quantifiers around such match rules,
the ones corresponding to the universal quantifiers are implicit in the concrete
syntax while
the ones corresponding to the $\nabla$-quantifiers  are explicit.
Our design for \mlts places the implicit quantifiers at outermost
scope: that is, the quantification over a match rule is of the form
$\forall\nabla$.
Another choice might be to allow some (all) universal quantifiers to
be explicitly written and placed among any \lsti{nab} bindings.
While this is a sensible choice, the $\forall\nabla$-prefixes is, in
fact, a reduction class in the sense that if one has a $\forall$
quantifier inside a $\nabla$-quantifier, it is possible to rotate
that $\nabla$-quantifier inside using a technique called
\emph{raising}~\cite{miller91jlc,miller05tocl}.
That is, the formula $\nabla x:\gamma\forall y:\tau (B x y)$ is
logically equivalent to the formula
$\forall h:(\gamma\ra\tau)\nabla x:\gamma (B x (h x))$: note that as
the $\nabla$-quantifier of type $\gamma$ is moved to the right over a
universal quantifier, the type of that quantifier is raised from $\tau$
to $\gamma\ra\tau$.
Thus, it is possible for an arbitrary mixing of $\forall$ and $\nabla$
quantifiers to be simplified to be of the form $\forall\nabla$.

\subsection{Nominal abstraction}

Before we can present the formal operational semantics of \mlts, we
need to introduce one final logical concept, \emph{nominal
  abstraction}, which allows implicit bindings represented by nominals
to be moved into explicit abstractions over terms~\cite{gacek11ic}.
The following notation is useful for defining this relationship.

Let $t$ be a term, let $c_1,\ldots,c_n$ be distinct nominals that
possibly occur in $t$, and let $y_1,\ldots,y_n$ be distinct variables
not occurring in $t$ and such that, for $1 \leq i \leq n$, $y_i$ and
$c_i$ have the same type. Then we write $\lambda c_1
\ldots\lambda c_n. t$ to denote the term $\lambda y_1 \ldots \lambda
y_n . t'$ where $t'$ is the term obtained from $t$ by replacing
$c_i$ by $y_i$ for $1\leq i\leq n$.
There is an ambiguity in this notation in that the choice of variables
$y_1,\ldots,y_n$ is not fixed.
This ambiguity is, however, harmless since the terms that are produced by
acceptable choices are all equivalent under $\alpha$-conversion.

Let $n\ge 0$ and let $s$ and $t$ be terms of type $\tau_1 \to \cdots
\to \tau_n \to \tau$ and $\tau$, respectively; notice, in particular,
that $s$ takes $n$ arguments to yield a term of the same type as $t$.
The formula $s \unrhd t$ is a \emph{nominal abstraction of degree $n$}
(or, simply, a \emph{nominal abstraction}).
The symbol $\unrhd$ is overloaded since it can be use at different
degrees (generally, the degree can be determined from context).
The nominal abstraction $s \unrhd t$ of degree $n$ is said to hold
just in the case that $s$ is $\lambda$-convertible to $\lambda
c_1\ldots c_n.t$ for some distinct nominals $c_1,\ldots,c_n$.

Clearly, nominal abstraction of degree $0$ is the same as equality
between terms based on $\lambda$-conversion, and we will use
$=$ to denote this relation in that case.
In the more general case, the term on the left of the operator serves
as a pattern for isolating occurrences of nominals.
For example, if $p$ is a binary constructor and $c_1$ and $c_2$
are nominals, then the nominal abstractions of the first row below
hold while those in the second row do not.
 \begin{align*}
\lambda x. x &\unrhd c_1 &
\lambda x. p\ x\ c_2 &\unrhd p\ c_1\ c_2 &
\lambda x. \lambda y. p\ x\ y &\unrhd p\ c_1\ c_2 \\
\lambda x. x &\not\mathrel\unrhd p\ c_1\ c_2 &
\lambda x. p\ x\ c_2 &\not\mathrel\unrhd p\ c_2\ c_1 &
\lambda x. \lambda y. p\ x\ y &\not\mathrel\unrhd p\ c_1\ c_1
\end{align*}

A logic with equality generalized to nominal abstraction has been
studied in~\cite{gacek09phd,gacek11ic} where a logic, named $\Gee$,
that contains fixed points, induction, coinduction,
$\nabla$-quantification, and nominal abstraction is given a sequent
calculus presentation.
Cut-elimination for $\Gee$ is proved in~\cite{gacek09phd,gacek11ic} and
algorithms and implementations for nominal abstraction are presented
in~\cite{gacek09phd,wang13ppdp}.
An important feature of the Abella prover---$\nabla$ in the head of a
definition---can be explained and encoded using nominal
abstraction~\cite{gacek08lics}.

\subsection{Natural semantics specification of \mlts}
\label{sec:natsem}

\begin{figure*}[t]
\begin{mathpar}
    \begin{array}{lcl}
    & & \text{values}
      \\
    V
    & \gramdef
    & X
    \\
    & \mid
    & \lam (\lambda x. M~x)
    \\
    & \mid
    & \back (\lambda X. V~X)
    \\
    & \mid
    & \hbox{variant}~c~[V_1,\dots,V_n]
  \end{array}

  \infer{\vdash\eval{\lam~R}{\lam~R}}{}

  \infer{\vdash\eval{\hbox{variant}~c~[T_1,\dots,T_n]}
                    {\hbox{variant}~c~[V_1,\dots,V_n]}}
        {\vdash \forall i \in [1; n],\;\eval{T_i}{V_i}}

  \infer{\vdash\eval{\new{(\lambda X.E\,X)}}{V}}
        {\vdash\nabla X.\eval{(E~X)}{V}}

  %
  %
  \infer{\vdash\eval{\hbox{app}~M~N}{V}}
        {\vdash\eval{M}{\lam~R}\quad
         \vdash\eval{N}{U}\quad
         \vdash\eval{(R~U)}{V}}

  \infer{\vdash\eval{(\Let{M}{R})}{V}}
        {\vdash\eval{M}{U}\quad\vdash\eval{(R~U)}{V}}

  \infer{\vdash\eval{\fix{R}}{V}}
        {\vdash\eval{R~(\fix{R})}{V}}

  \infer{\vdash\eval{\arob{M}{X}}{V}}
        {\vdash\eval{M}{\back R}\quad
         \vdash\eval{(R~X)}{V}}

  \infer{\vdash\eval{\back {(\lambda X.E\,X)}}
                    {\back {(\lambda X.V\,X)}}}
        {\vdash\nabla X.\eval{(E~X)}{(V~X)}}

  \infer{\vdash\eval{(\match{T}{(\cons{Rule}{Rules})})}{V}}
        {\vdash\clause{T}{Rule}{U}\quad\vdash\eval{U}{V}}

  \infer{\vdash\eval{(\match{T}{(\cons{Rule}{Rules})})}{V}}
        {\vdash \neg(\exists u,\,\clause{T}{Rule}{u})\quad
         \vdash\eval{(\match{T}{Rules})}{V}}

\infer{\vdash\clause{T}{(\hbox{all}~{(\lambda x. P~x)})}{U}}
      {\vdash\exists x .\clause{T}{(P~x)}{U}}

\infer{\vdash\clause{T}{(\hbox{nab}~Z_1\ldots\hbox{nab}~Z_m.(p \Longrightarrow u))}{U}}
      {\vdash\matches{T}{P}\quad
       \vdash (\lambda Z_1\ldots\lambda Z_m.(p \Longrightarrow u))\unrhd
                                            (P \Longrightarrow U)}

  \infer{\vdash\matches{(\hbox{variant}~c~[t_1,\dots,t_n])}
                       {(\hbox{pvariant}~c~[p_1,\dots,p_n])}}
        {\vdash \forall i \in [1; n],\, \matches{t_i}{p_i}}

  \infer{\vdash \matches{c}{(\hbox{pnom}~c)}}{\hbox{nominal}(c)}

  \infer{\vdash \matches{x}{(\hbox{pvar}~x)}}{}
\end{mathpar}
\caption{A natural semantics specification of evaluation.}%
\label{fig:natsem}
\end{figure*}

We can now define the operational semantics of \mlts by giving
inference rules in the style of natural semantics (a.k.a.\ big-step
semantics) following Kahn~\cite{kahn87stacs}.
The semantic definition for the core of \mlts is defined in
Figure~\ref{fig:natsem}.
Since those inference rules are written using a higher-order abstract
syntax for \mlts, directly inspired by \lP term representations; we
briefly describe how that abstract syntax is derived from the concrete
syntax.

Instead of detailing the translation from concrete to abstract syntax,
we illustrate this translation with an example.
There is an implementation of \mlts that includes a parser and a
transpiler into \lP code: this system is available for
online use and for download at \url{https://trymlts.github.io}
\cite{trymlts}.
For example, the \lP code in Figure~\ref{fig:abs size} is the abstract
syntax for the  \mlts program for \lsti{size} given in
Section~\ref{sec:examples}.

\begin{figure}[t]
\begin{lstlisting}[language=lprolog]
(fix size \ lam term \
  match term
  [(all m \ all n \
     (pvariant c_App [(pvar n), (pvar m)]) ==>
     (special add [(special add [(int 1),
                   (app size n)]),
                   (app size m)])),
   (all r \ (pvariant c_Abs [pvar r]) ==>
      (special add
               [(int 1),
                (new X \ app size
                             (arobase r X))])),
   (nab X \ (pnom X) ==> (int 1))])
\end{lstlisting}

\caption{The abstract syntax of the \lsti{size} program.}
\label{fig:abs size}
\end{figure}

The backslash (as infix notation) is also used in \lP to denote
binders and it is the only $\lambda$Prolog primitive in
Figure~\ref{fig:abs size}. The other constructors are introduced to
encode \mlts abstract syntax trees.

This encoding of \mlts syntax is a generalization of the
familiar semantic encoding of the \ulc given by Scott in
1970~\cite{scott70}, in which a semantic domain $D$ and two continuous
mappings (retracts) $\Phi\colon D \ra (D \ra D)$ (encoding
application) and $\Psi\colon (D \ra D) \ra D$ (encoding abstraction)
are used to encode the \ulc.
For example, the \ulc $\lambda x\lambda y ((x y) y)$
is encoded as a value in domain $D$ using the expression
$(\Psi(\lambda x(\Psi(\lambda y(\Phi (\Phi~x~y)~y))))))$.
In Figure~\ref{fig:abs size}, the constructors \lsti{c_App} and
\lsti{c_Abs} represents the $\Phi$ and $\Psi$ functions,
respectively.
The \lP abstraction operator (backslash) is used to build
expressions that correspond to inhabitants of $D\ra D$.

The constant \lstip{fix} represents anonymous fixpoints, to which recursive
functions are translated (we also have an n-ary fixpoint for
mutually-recursive functions).
Note that \lstip{fix x \ t} is idiomatic \lP{} syntax for
the application \lstip{fix (x \ t)}, omitting parentheses to use
\lstip{fix} in the style of a binder.

The expression \lstip{lam x \ ...} represents the \mlts expression
\lsti{fun x -> ...}; in our abstract syntax we write
$\lam {(\lambda X. \dots)}$.
(We do not make a syntactic distinction between $X$ and $x$ which are
just variables, but we use uppercase variables in the abstract syntax
for variables that represent nominals in the language.)
Similarly, the expression \lstip{new X \ ...} encodes \lsti{new X in ...}; in
our abstract syntax we write $\new {(\lambda X. \dots)}$.
The expression-former \lstip{match} represents pattern-matching, it
expects a scrutinee and a list of clauses.
Clauses are built from the infix operator \lstip{==>}, taking
a pattern on the left and a term on the right, and from quantifiers
\lstip{all}, to introduce universally-quantified variables
(implicit in \mlts programs), and \lstip{nab} to introduce
nominals. \lstip{all}-bound variables and \lstip{nab}-bound nominals
have the type of expressions; they are injected in patterns by
\lstip{pvar} and \lstip{pnom}.
\lstip{pvariant} (in patterns) and \lstip{variant} (in expressions)
denote datatype constructor applications, they expect a datatype
constructor and a list of arguments.
\lstip{special} expects the name of a run-time primitive
(arithmetic operations, polymorphic equality...) and a list of
arguments. \lstip{int} represents integer literals.
Finally, we use explicit AST expression-formers \lstip{backslash} and
\lstip{arobase} (a French name for \lsti{@}) and pattern-formers
\lstip{pbackslash} and \lstip{parobase} to represent the constructions
\lsti{\} and \lsti{@} of \mlts. Only \lstip{arobase} is present in
  this example.

It is intended that the inference rules given in
Figure~\ref{fig:natsem} are, in fact, notations for formulas in the
logic $\Gee$.
For example, schema variables of the inference rules are universally
quantified around the intended formula; the horizontal line is an
implication; the list of premises is a conjunction; and $\Downarrow$
is a binary (infix) predicate, etc.
Some features of $\Gee$ are exploited by some of those
inference rules: those features are enumerated below.

Figure~\ref{fig:natsem} starts with a grammar for values. In addition
to lambda-abstractions, $\back$-expressions (with a value as the body)
and variant values, (open) values also include nominals. Evaluating
a closed term can never produce a nominal, but evaluation rules under
binders may return nominals.

In the rules for \lstip{app}, \lstip{let} and \lstip{fix}, a variable
of arity type $\bfz\ra\bfz$ (namely, $R$) is applied to a term of
arity type $\bfz$.
These rules make use of the underlying equality theory of simply typed
$\lambda$-terms in $\Gee$ to perform a substitution.
In the rule for apply, for example, if $R$ is instantiated to the term
$\lambda w. t$ and $U$ is instantiated by the term $s$, then the
expression written as $(R~U)$ is equal (in $\Gee$) to the result of
substituting $s$ for the free occurrences of $w$ in $t$: that is, to
the result of a $\beta$-reduction on the expression
$((\lambda w. t)~s)$.
(While matching and applying patterns is limited to
$\beta_0$-reduction, full $\beta$-reduction is used for the natural
semantic specification.)

Existential quantification is written explicitly into the first rule
for patterns. We write it explicitly here to highlight the fact that
solving the problem of finding instances of pattern variables in
matching rules is lifted to the general problem of finding
substitution terms in $\Gee$.


The proof rules for natural semantics are nondeterministic in
principle.
Consider attempting to prove that $t$, a term of arity
type $\bfz$, has a value: that is, $\exists V, \eval{t}{V}$.
It can be the case that no proof exists or that there might be several
proofs with different values for $V$.
No proofs are possible if, for example, the condition in a conditional
phrase does not evaluate to a boolean or if there are insufficient
match rules provided to cover all the possible values given to a match
expression.
Ultimately, we will want to provide a static check that could issue
a warning if the rules listed in a match
expression are not exhaustive.
Conversely, the variables introduced by \lstip{all} and \lstip{nab} in
patterns may have several satisfying values, if they are not used in
the pattern itself, or only in flexible occurrences
(see~Section~\ref{subsubsec:rigid-occurrence}).
%


The \emph{nominal abstraction} of $\Gee$ is directly invoked to solve
pattern matching in which nominals are explicitly abstracted using the
\lsti{nab} binding construction.
When attempting to prove the judgment \(\vdash\clause{T}{Rule}{U}\), the
inference rules in Figure~\ref{fig:natsem} eventually lead to an
attempt to prove in $\Gee$ an existentially quantified nominal
abstraction of the form
\[
  \exists x_1\ldots\exists x_n[
     (\lambda Z_1\ldots\lambda Z_m. (p \Longrightarrow u)) \unrhd
                                    (P \Longrightarrow U)].
\]
Here, the arrow $\Longrightarrow$ is simply a formal (syntactic)
pairing operator, expecting a pattern on the left and a term on the right.
The schema variables $x_1,\ldots, x_n$ can appear free only in $p$ and
$u$.


The last ingredient of our pattern-matching rule is the judgment
$(\vdash\matches{T}{P})$ that checks that a term or value
$T$ is indeed matched by a pattern $P$.
%
%
Since patterns and terms are encoded using two
distinct syntactic categories, this judgment relates pattern-formers
to the corresponding term-formers. Nominals are embedded in patterns
by the $\hbox{pnom}(c)$ pattern-former, which matches a corresponding
nominal---the condition $\hbox{nominal}(c)$ can be expressed in terms
of nominal abstraction $(\lambda X.\,X) \unrhd c$. Term variables
introduced by $\hbox{all}$ are embedded in patterns by the
$\hbox{pvar}$ pattern-former, and they can match any term $x$---note
that in this rule, $x$ denotes an arbitrary term, substituted for
a term variable by the $\hbox{all}$-handling rule.


It is worth pointing out that given the way we have defined the
operational semantics of \mlts, it is immediate that ``nominals cannot
escape their scopes.''
For example, the expression \lsti{(new X in X)} does not have a
value (in abstract syntax, this expression translates to $\new{(\lambda X. X)}$).
More precisely, there is no proof of
$\vdash\exists v. \eval{(\new{(\lambda X. X)})}{v}$
using the rules in Figure~\ref{fig:natsem}.
To see why this is an immediate consequence of the specification
of evaluation, consider the formula (which encodes the rule
in Figure~\ref{fig:natsem} for \lstip{new})
\[
\forall E\forall V[(\nabla X.\eval{(E~X)}{V})\supset (\eval{\new{E}}{V})].
\]
Given that the scope of the $\nabla X$ is inside the scope of $\forall
V$, it is not possible for any instance of this formula to allow the
$X$ binder to appear as the second argument of the $\Downarrow$
predicate.
While such escaping is easily ruled out using this logical
specification, a direct implementation of this logic may incur a
cost, however, to constantly ensure that no escaping is permitted.
(See Section~\ref{sec:escape} for more discussion on this point.)

\begin{figure*}[t]
\[
  \infer{\judty{\Gamma,x:C}{x}{C}}{\strut}
  \quad
  \infer{\judty{\Gamma}{(M N)}{B}}
        {\judty{\Gamma}{M}{A -> B} \qquad \judty{\Gamma}{N}{A}}
  \quad
  \infer{\judty{\Gamma}{(fun x -> M)}{A -> B}}
        {\judty{\Gamma, \lsti{x}:\lsti{A}}{M}{B}}
  \qquad
  \infer{\judty{\Gamma}{let x = M in N}{B}}
        {\judty{\Gamma}{M}{A} \qquad \judty{\Gamma,\lsti{x}:\lsti{A} }{M}{B}}
  \qquad
  \infer{\judty{\Gamma}{(X \\ M)}{A => B}}
        {\judty{\Gamma, \lsti{X}:\lsti{A}}{M}{B} \qquad \openty{A}}
\]

\[
  \infer{\judty{\Gamma}{(r @ t1 ... tn)}{A}}
        {\judty{\Gamma}{r}{A1 => ... => An => A}\quad
         \judty{\Gamma}{t1}{A1}\quad\ldots\quad\judty{\Gamma}{tn}{An}}
\qquad
\infer{\judty{\Gamma}{C(t1,...,tn)}{B}}
      {\lsti{C} : \lsti{A1}, \ldots, \lsti{An} \to \lsti{B}
       \qquad
       \judty{\Gamma}{t1}{A1}
       \quad\ldots\quad
       \judty{\Gamma}{tn}{An}}
\]

\[
  \infer{\judty{\Gamma}{(M, N)}{A * B}}
        {\judty{\Gamma}{M}{A}%
         \quad%
         \judty{\Gamma}{N}{B}}
  \qquad
  \infer{\judty{\Gamma}{(new X in M)}{B}}
        {\judty{\Gamma, \lsti{X}:\lsti{A}}{M}{B} \qquad \openty{A}}
  \quad
  \infer{\judty{\Gamma}{match term with R1 | ... | Rn}{A}}
        {\judty{\Gamma}{term}{B} \qquad
         \judr{\Gamma}{B}{R1}{A} \qquad \ldots \qquad \judr{\Gamma}{B}{Rn}{A}}
\]

\[
  \infer{\judr{\Gamma}{A}{nab X in R}{B}}
        {\judr{\Gamma,\lsti{X}:\lsti{C}}{A}{R}{B} \qquad \openty{C}}
  \qquad
  \infer{\judr{\Gamma}{A}{L -> R}{B}}
        {\judlh{\Gamma}{L}{A}{\Delta} \qquad \judty{\Gamma,\Delta}{R}{B}}
  \qquad
  \infer{\judlh{\Gamma}{(r @ X1 ... Xn)}{A}
               {\lsti{r}:\lsti{A1 => ... => An => A}}}
        {\judty{\Gamma}{X1}{A1}~\ldots~\judty{\Gamma}{Xn}{An}\quad
         \openty{A1}\ldots\openty{An}}
\]

\[
  \infer{\judlh{\Gamma}{x}{A}{\lsti{x}:\lsti{A}}}{\strut}
  \qquad
  \infer{\judlh{\Gamma}{(p,q)}{A * B}{\Delta_1,\Delta_2}}
        {\judlh{\Gamma}{p}{A}{\Delta_1}\qquad \judlh{\Gamma}{q}{B}{\Delta_2}}
  \qquad
  \infer{\judlh{\Gamma}{C(p1,...,pn)}{B}{\Delta_1,\ldots,\Delta_n}}
        {\lsti{C} : \lsti{A1}, \ldots, \lsti{An} \to \lsti{B}
         \qquad
         \judlh{\Gamma}{p1}{A1}{\Delta_1}
         \quad\ldots\quad
         \judlh{\Gamma}{pn}{An}{\Delta_n}}
\]

\caption{Typing rules based on the concrete syntax for the new
  features of \mlts.}
\label{fig:typing}
  \end{figure*}

\section{Typing rules and restrictions, small-step semantics, meta-theorems}%
\label{sec:formal}
In this section we present a typing discipline for \mlts, followed by a few restrictions
on pattern matching necessary for it to remain well behaved and the establishment of
standard formal results.

\subsection{Typing}\label{sec:typing}

Given that \mlts is a rather mild extension of OCaml at the syntax
level, a typing system for \mlts is simple to present and follows
standard practices.
Figure~\ref{fig:typing} contains the rules for typing the new features
of \mlts: additional rules for encoding \lsti{let} and \lsti{let rec}
constructions (as well as for built-in types such as integers) must
also be added, but these follow the usual pattern.
The inference rules in this figure involve the following typing
judgments.
\[
\judty{\Gamma}{M}{A}
\qquad
  \judr{\Gamma}{A}{R}{B}
\qquad
\judlh{\Gamma}{M}{A}{\Delta}
\qquad
\openty{A}
  \]
In all of these rules, $\Gamma$ is the usual association between bound
variables and a type: in our situation, $\Gamma$ will associate both
variables and nominals to type expressions.
(We also assume that the order of pairs in $\Gamma$ is not important.)
%
The first of these judgments is the usual typing judgment between a
program expression \lsti{M} and \lsti{A}.
The second of these judgments is used to type a clause
\lsti{R} that has a left-hand side of type \lsti{A} and a right-hand
side of type \lsti{B}.
For example, the following typing judgment should be provable.
\[
\Gamma \vdash \lsti{tm} :
              \lsti{Abs(r) -> 1 + (new X in size (r @ X))} :
              \lsti{int}
\]
Since this rule expression is intended to be closed (that is, the
variable \lsti{r} is quantified implicitly around this rule), the
actual value of $\Gamma$ will not impact this particular typing
judgment.
The third typing judgment above is used to analyze the left-hand-side
of a match rule: in particular,
$\judlh{\Gamma}{M}{A}{\Delta}$ holds if during the process of analyzing
the pattern \lsti{M}, pattern variables are produced (since these are
implicitly quantified) and placed into the typing context $\Delta$.
For example, the following should be provable.
\[
\Gamma \vdash \lsti{Abs(r)} : \lsti{tm} \dashv
              \{ \lsti{r} : \lsti{tm => tm} \}
\]

Some of the inference rules in Figure~\ref{fig:typing} contain
premises of the form $(\openty{A})$ where \lsti{A} is a primitive type.
Types for which this judgment holds are called \emph{open types} and are the types of
bindings in the \lsti{new} and backslash expressions: equivalently,
open types can contain nominals.
For our purposes here, we can assume that every type that is defined
in a program (using the \lsti{type} command) is presumed to be open.
For example, the judgment $(\openty{tm})$ needs to be true so that the
type \lsti{tm => tm} can be formed in the various typing rules.
On the other hand, the built-in type for integers \lsti{int} should
not be considered open in this sense.
Clearly a keyword must be added to datatype declarations to indicate
if a type is intended as open in this sense.

In the inference rules in Figure~\ref{fig:typing}, whenever we extend
the typing context $\Gamma$ to, say, $\Gamma, \lsti{X} : \lsti{A}$, we
 assume that \lsti{X} is not declared in $\Gamma$
already.
Since $\alpha$-conversion is always possible within terms, this
assumption can always be satisfied.
Note that since pattern variables are restricted (as is usual) so that
they have at most one occurrence in a given pattern, the union of
contexts, in the form $\Delta_1,\ldots,\Delta_n$ never attributes more
than one type to the same variable.

The prototype implementation TryMLTS~\cite{trymlts} of \mlts contains
a type inference engine that runs on top of \lP: given the
hypothetical judgments available in \lP, the implemented typing system
is structured differently (but equivalently) to the one given in
Figure~\ref{fig:typing}.
By using \lP, we were able to turn this typing system into one that
does type inference: this type inference engine does not infer
polymorphic typing, however.

\subsection{Restriction on matching}\label{ssec:unification}

Since we are not able to decompose bindings into their bound variable
and body, we need to find alternative means for analyzing the
structure of terms containing bindings.
As our earlier examples illustrated, matching within patterns can
be used to probe terms and their bindings.
If we do not place restrictions on the use of pattern variables, then
patterns can have complex behaviors that we may wish to avoid during
evaluation.

\subsubsection{Unique occurrence of pattern variables.}

We impose a familiar restriction on the match rules: a pattern
variable must have exactly one occurrence within a match pattern.
Asking for at least one occurrence avoids under-specified pattern variables,
that could be bound to anything.
As is typical in ML-style languages, asking for at most one occurrence
relieves pattern matching from the need to check equality of terms.
Since terms can be large, pattern matching could involve a costly recursive
descent of terms; we forbid repeated occurrences of pattern
variables and force the programmer to insert equality checking outside
the pattern matching operation.
Thus, instead of defining \lsti{memb : tm -> tm list -> bool} with the
following code using a repeated match variable
\begin{lstlisting}[language=mlts,linerange={5-8}]
let rec memb x l = match (x,l) with
    | (x,[])     -> false
    | (x,(x::l)) -> true
    | (y,(x::l)) -> memb x l;;
\end{lstlisting}
we can require the programmer to write an equality predicate for type
\lsti{tm} and then rewrite the program above as follows.
\begin{lstlisting}[language=mlts,linerange={10-21}]
let rec eqtm t s = match (t,s) with 
    | (App(m1,m2), App(n1,n2)) -> eqtm m1 n1 && 
                                  eqtm m2 n2
    | (Abs r, Abs s)  -> new X in eqtm (r @ X)
                                       (s @ X)
    | nab X in (X, X) -> true
    | _               -> false;;

let rec memb x l = match (x,l) with
    | (x,[])     -> false
    | (x,(y::l)) -> if (eqtm x y)
                    then true else (memb x l);;
\end{lstlisting}
Given the definition of the \lsti{tm} datatype, it is clear that a
compiler for \mlts could define its own equality predicate for this
type.
In that case, repeated variable occurrences in patterns could be
allowed since resolving such patterns could be done using these equality
predicates.

\subsubsection{Restricted use of higher-order pattern variables.}
\label{sssec:hop}

Since pattern variables within match rules can have higher-order
types, occurrences of those variables within patterns need to be
restricted: otherwise, undesirable features of higher-order matching
could appear.
Fortunately, there is a natural restriction on occurrences of pattern
variables that guarantees that a match either fails or succeeds with
at most one solution.
That restriction is the following: every occurrence of an expression
of the form \lsti{(r @ X1 ... Xn)} in the left-hand side of a match
rule must be such that the pattern variable \lsti{r} is applied to
$n\ge0$ \emph{distinct} nominals \lsti{X1 ... Xn} and those nominals are
bound \emph{within} the scope of the binding for \lsti{r}.
For example, the following expression is not well formed
\begin{lstlisting}[language=mlts,linerange={5-7}]
Abs(X\ (match Abs(Y\ App(X,Y)) with
            | Abs(Z\ r @ Z X) ->
                       Abs(Z\ r @ X Z)));;
\end{lstlisting}
since the scope of the nominal \lsti{X} contains
the (implicit) scope of the pattern variable \lsti{r}, which is
around the rule \lsti{(Abs(Z\ r @ Z X) -> Abs(Z\ r @ X Z))}.

This restriction can be motivated within a purely logical
setting as follows.
Let $j$ be a primitive type and let $F : j\ra j\ra j$ be a
simply typed constant.
The formula $\exists g:j\ra j~\forall X:j~[g~X = (F~X~X)]$ has a
unique proof in which $g$ is instantiated by the term $\lambda
W. (F~W~W)$.
Note that the binding scope of the variable $X$ is inside the binding scope
of the variable $g$.
If, however, one switches the order of the quantifiers, yielding
$\forall X:j~\exists g:j\ra j~[g~X = (F~X~X)]$, then there are four
different proofs of this equation: if one replaces the outermost
universal quantifier with an eigenvariable (or nominal), say $A$, then
there are four different solutions for $g$, namely,
$\lambda W. (F~A~A)$,
$\lambda W. (F~A~W)$,
$\lambda W. (F~W~A)$, and
$\lambda W. (F~W~W)$.

The subset of higher-order unification in which unification variables
(a.k.a., logic variables, meta-variables, pattern variables) are
applied to distinct bound variables restricted as described above, is
called \emph{higher-order pattern unification} or \Ll \emph{unification}
\cite{miller91jlc}.
(We assume here the usual convention that unification problems and
matching problems only involve terms that are in $\beta$-normal form.)
This particular subset of higher-order unification is commonly
implemented in theorem provers such as Abella~\cite{baelde14jfr},
Minlog~\cite{schwichtenberg06provers}, and Twelf~\cite{pfenning99cade}
as well as recent implementations of \lP
\cite{dunchev15lpar,teyjus.website}.
A functional programming implementation of such unification is given
in~\cite{nipkow93lics}.

The following results about higher-order pattern unification are
proved in~\cite{miller91jlc}.
\begin{enumerate}
\item It is decidable and unitary, meaning that if there is a unifier then there
  exists a most general unifier.
\item It does not depend on typing.  As a result, it is
  possible to add it to the evaluator for \mlts based on untyped
  terms.
\item The only form of $\beta$-conversion that is needed to solve such
  unification problems is what is called $\beta_0$-conversion which is
  a form of the $\beta$ rule that equates $(\lambda x.t)x$ with $t$.
\end{enumerate}

An equivalent way to write the $\beta_0$-conversion rule (assuming the
presence of $\alpha$-conversion) is that $(\lambda x.t)y$ converts to
$t[y/x]$ \emph{provided} that $y$ is not free in $\lambda x.t$.
Notice that applying $\beta_0$ reduction actually makes a term
smaller and does not introduce new $\beta$ redexes: as a result it is
not a surprise that such unification (and, hence, matching) has low
computational complexity.


\subsubsection{All \lsti{nab} bound variables must have a rigid occurrence.}
\label{subsubsec:rigid-occurrence}

\begin{figure}[b]
\vspace{-1em}
\begin{lstlisting}[language=mlts,linerange={9-12}]
Abs(X\ (match Abs(Y\ App(X,Y)) with 
            | nab W in Abs(Z\ r @ Z W) -> 
                       Abs(Z\ r @ W Z)));;
\end{lstlisting}
\caption{Code that does not satisfy the restriction on occurrences of
  \lsti{nab} bound variables.}
\label{fig:flex}
\end{figure}

There is an additional restriction on match rules that is associated
to the \lsti{nab} binder that appear in such rules.
We say that an occurrence of a \lsti{nab}-quantified nominal is
\emph{flexible} if it is in the scope of an \lsti{@}.
For example, in the code in Figure~\ref{fig:flex},
the nominal binding \lsti{W} has two occurrences that are flexible: one
each within \lsti{(r @ Z W)} and \lsti{(r @ W Z)}.
All other occurrences of a \lsti{nab}-bound nominal are
\emph{rigid}.  For example, in the match rule
\lsti{| nab X in X -> 1}, \lsti{X} has a binding occurrence and a
rigid occurrence.  In the auxiliary function used by the
\lsti{index} function in Figure~\ref{fig:deb}, namely,
\begin{lstlisting}
let rec aux c x k = match (x, k) with
    | nab X   in (X, X::(l @ X))   -> c
    | nab X Y in (X, Y::(l @ X Y)) ->
                        aux (c + 1) x (l @ X Y)
\end{lstlisting}
the nominals \lsti{X} and \lsti{Y} have both rigid and flexible
occurrences within their scope.

The one additional restriction that we need is the following: every
\lsti{nab}-bound variable must have at least one rigid occurrence in
the left part of the match rule (the pattern) that falls within the
scope of its binder.
For example, the code in Figure~\ref{fig:flex} does not satisfy this
restriction since every occurrence of \lsti{W} in the pattern is
flexible (there is just one such occurrence).

This restriction ensures that each \lsti{nab}-bound nominal in
a matching clause is mapped to a uniquely-determined nominal of the
ambient context.
As interesting counter-examples, consider
\begin{lstlisting}
  match Z with
  | nab X Y in  (r @ X Y) -> term
\end{lstlisting}
where \lsti{Z} is a nominal, and
\begin{lstlisting}
  match 1 with
  | nab X in 1 -> t
\end{lstlisting}
which are both ruled out by this restriction.
In the first example, there are two instantiations for \lsti{r} that
make this match succeed, namely, using the terms \lsti{X\Y\X} and
\lsti{X\Y\Y}. This breaks the determinacy property --
Theorem~\ref{thm:determinacy}.
In the second example, the nominal \lsti{X} is completely unconstrained by the
pattern.
If this program was allowed, our natural semantics dictates that it
should behave as \lsti{new X in t}; the restriction guarantees that
\lsti{new} is the only language construct that may introduce dynamic
nominal-escape failures.

\subsection{Small-step operational semantics}

\begin{mathparfig}{fig:small-step-core}{Small step reduction: core fragment}
  \begin{array}{lcl}
    & & \text{evaluation contexts}
      \\
    E[\hole]
    & \gramdef
    & \hole
    \\
    & \mid
    & \hbox{app}~M~E
      \mid
      \hbox{app}~E~N
    \\
    & \mid
    & \back (\lambda X. E)
      \mid
      \arob E X
    \\
    & \mid
    & \new (\lambda x. E)
    \\
    & \mid
    & \hbox{variant}~c~[ M_1 \dots M_k, E, M_{k+2} \dots M_n ]
    \\
    & \mid
    & \match E {[ R_1, \dots, R_n ]}
  \end{array}

  \infer
  {\rtohd {\hbox{app}~(\lam~R)~V} {R~V}}
  { }

  \infer
  {\rtohd {\arob {(\back R)} X} {R~X}}
  { }

  \infer
  {\rtohd {\fix R} {R~(\fix{R})}}
  { }

  \infer
  {\rto {E[M]} {E[M']}}
  {\rtohd M {M'}}

  \infer
  {\rto {E[\new (\lambda X.V)]} {E[V]}}
  {X \notin V}
\end{mathparfig}

\begin{mathparfig}{fig:paths}{Rigid paths in values and patterns}
    \begin{array}{lcl}
      & & \text{rigid paths}
      \\
      \pi
      & \gramdef
      & \hole
      \\
      & \mid & \hbox{variant}~C~i~\pi
      \\
      & \mid & \back (\lambda X. \pi)
      \\
      & \mid & \arob \pi X \\
    \end{array}
\\
    \fbox{Rigid occurrence in a value: $\atpath {V'} \pi V$}
\\
    \infer
    { }
    {\atpath {V'} \hole {V'}}

    \infer
    {\atpath {V'} {(\hbox{variant}~C~k~\pi)} {\hbox{variant}~c~[V_1, \dots, V_n]}}
    {\atpath {V'} \pi {V_k}}

    \infer
    {\atpath {V'} {(\back {(\lambda X. \pi)})} {(\back {(\lambda X. V)})}}
    {\nabla X.\;\atpath {V'} \pi V}

    \infer
    {\atpath {V'} {(\arob \pi X)} {V}}
    {\atpath {V'} \pi {\back {(\lambda X. V)}}}
\\
    \fbox{Rigid occurrence in a pattern: $\atpath {p'} \pi p$}
\\
    \infer
    { }
    {\atpath {p'} \hole {p'}}

    \infer
    {\atpath {p'} {(\hbox{variant}~C~k~\pi)} {\hbox{variant}~c~[p_1, \dots, p_n]}}
    {\atpath {p'} \pi {p_k}}

    \infer
    {\atpath {p'} {(\back {(\lambda X. \pi)})} {(\back {(\lambda X. p)})}}
    {\nabla X.\;\atpath {p'} \pi p}

    \infer
    {\atpath {p'} {(\arob \pi X)} {(\arob p X)}}
    {\atpath {p'} \pi p}
\\
    \fbox{Rigid occurrence in a clause: $\atpath {p'} \pi R$}
\\
    \infer
    {\atpath {p'} \pi {\nab (\lambda Z. R)}}
    {\nabla Z.\;\atpath {p'} \pi R}

    \infer%
    {\atpath {p'} \pi {\hbox{all}~R}}%
    {\nabla x.\;\atpath {p'} \pi {R~x}}

    \infer
    {\atpath {p'} \pi {\tyto p M}}
    {\atpath {p'} \pi p}
\end{mathparfig}

\begin{mathparfig}{fig:small-step-patterns}{Small step reduction: pattern-matching}
  \infer{\rtohd{\match{V}{(\cons{R}{Rs})}} N}
        {\predwith V {R} \emptyset N}

  \infer{\rtohd{\match V {(\cons{R}{Rs})} }
               {\match V {Rs}}}
  {\not \exists N.\, (\predwith V {R} \emptyset N)}

  \fbox{Matching a value against a clause: $\predwith V R \sigma N$}

  \infer{\predwith V {\nab R} \sigma N}
        {\nabla X.\,\exists \pi, Y.
         \quad
         \begin{array}{l}
           \atpath X \pi R~X
           \\
           \atpath Y \pi V
         \end{array}
         \quad
         \begin{array}{l}
         Y \notin R~X \hfill Y \notin \sigma
         \\
         \predwith V {R~Y} \sigma N
         \end{array}}

  \infer{\predwith V {(\all R)} \sigma {N~{V_x}}}
        {\nabla x.\;\predwith V R {\plug \sigma {\sto x {V_x}}} {N~x}}

  \infer{\predwith V {(\ubranch p N)} \sigma N}
        {\predmatches V p \sigma}

  \fbox{Matching a value against a pattern: $\predmatches V p \sigma$}

  \infer{\predmatches {\back V} {\back p} \sigma}
        {\nabla X.\;\predmatches {V~X} {p~X} \sigma}

  \infer{\predmatches
            {V~X}
            {\arob p X}
            {\sigma}
        }
        {\predmatches {\back V} p \sigma}

  \infer{\predmatches {\ucons C {V_1  \dots V_n}}
                      {\ucons C {p_1  \dots p_n}}
                      {\biguplus_i \sigma_i}}
        {\forall i \in 1 .. n,
        \quad \predmatches {V_i} {p_i} {\sigma_i}}
  \\
  \infer{\predmatches V x { (\sto x V)}}{}
  \quad
  \infer{\predmatches V \wild \emptyset}{}
  \quad
  \infer{\predmatches X X \emptyset}{}
\end{mathparfig}

As a complement to the natural (big-step) semantics of
Figure~\ref{fig:natsem}, we developed a small-step operational
semantics of \mlts{}. Its two salient features are as follows: (1) the
small-step treatment of evaluation contexts clarifies the moments
during reduction where escape-checking must be performed (this is
often left implicit in the natural semantics), and (2) its treatment
of pattern-matching does \emph{not} use nominal-abstraction -- it
implements an equivalent but lower-level mechanism. This lower-level
expression of the handling of nabla-bound nominals in pattern-matching
gives a more operational intuition of the language, and it also guides
practical implementations in languages without native support for
nominal abstraction. In fact, we co-evolved this operational semantics
with the \lP implementation of the language, the former guiding the
latter, with the bugs found playing with the latter informing changes
to the former -- using the natural semantics as a reference
specification for what the behavior should be.

Due to space restrictions, we will not give a fully detailed
explanation of this operational semantics. For the details, the
figures will have to speak for themselves, we will below give
a high-level presentation of the rules.

\paragraph{Core language (without pattern-matching)}
Figure~\ref{fig:small-step-core} gives a small-step operational
semantics for the fragment of the language without
pattern-matching. We use the standard approach of decomposing
reduction into a head reduction and evaluation contexts.

Our evaluation contexts allow reduction under the nominal abstraction
($\back {(\lambda X. E)}$ is an evaluation context): it does not
delay computation like the standard $\lambda$-abstraction does.

The other non-standard aspect of this fragment is the treatment of the
name-creation construct $\new{(\lambda X. M)}$. Instead of trying to
``generate a fresh nominal'' in the small-step semantics, we simply
allow reduction under $\hbox{new}$ binders -- the stack of
$\hbox{new}$ in the current evaluation context is the set of ``ambient
nominals'' available at this point of the program execution. In
addition to the standard rule allowing reduction under context, we
have an extra contextual rule to allow popping a $\hbox{new}$ binder
off the context: when the term inside that binder has been fully
evaluated to a value, so we have a term of the form
$E[\new{(\lambda X. V)}]$, we can remove the binder after performing
an escape check ($X \notin V$), continuing evaluation with $E[V]$. If
the escape check fails, the term is stuck -- this is the presentation
in our semantics of nominal escape as a dynamic failure.

\paragraph{Paths of rigid occurrences} As we explained in
Section~\ref{subsubsec:rigid-occurrence}, a clause of the form
$\nab {(\lambda X. \ubranch p M})$ is only accepted if the nominal $X$
has at least one rigid occurrence in the pattern $p$. The operational
semantics uses this criterion. In Figure~\ref{fig:paths}, we define
a grammar of \emph{rigid paths} $\pi$, which represent evidence that
a given occurrence of a sub-pattern (sub-value) in a pattern (value)
is in rigid position, as defined by the judgments $\atpath {p'} \pi p$
and $\atpath {v'} \pi v$.

Looking at the path $(\arob \pi X)$ in a pattern
$(\arob p X)$ selects a sub-value by looking at $\pi$ in $p$. In terms,
$(\arob v X)$ is not a value, but any value $V~X$ can be eta-expanded
to the (non-value) form $(\arob {(\back {\lambda X. V~X})} X)$, so we
look for the sub-value at path $\pi$ in $(\back {\lambda X. V~X})$.

\paragraph{Operational semantics of pattern matching} The treatment of
pattern-matching in this operational semantics, given in
Figure~\ref{fig:small-step-patterns} is not particularly small-step:
matching a value against a clause is a single step, so it is more
big-step in nature. The key interest of these rules is that they do
not use nominal abstraction, and instead ``implement'' the same
behavior in a more computational style.

The judgment $(\predmatches v p \sigma)$ holds when the value $v$ can
be matched against the value $p$, by performing the substitution
$\sigma$ -- from pattern variables in $p$ into sub-values of $v$. The
inputs of the judgment are $v$ and $p$, and the substitution $\sigma$
is an output of the inference process.

The judgment $(\predwith v R \emptyset N)$ holds when the value $v$
can be matched against the clause $R$, returning a right-hand-side $N$
to evaluate. In $N$, the pattern variables bound in $R$ (by the
clause-former $\all {(\lambda x. R)}$) have already been substituted
with the corresponding sub-values of $v$. In the general case, we want
to define the meaning of matching a value $v$ against a clause $R$
\emph{after} having traversed some all-quantifications, that is with
extra pattern variables in the ambient context; the general form of
the judgment is $\predwith v R \sigma N$, where $\sigma$ is
a substitution from those ambient pattern variables, which still occur
free in $N$.

The correspondence with the natural semantics is as follows:
$\predwith v R \sigma N$ in the operational semantics holds if and only
if $\clause{v}{R[\sigma]}{N[\sigma]}$ holds in the natural semantics.

\subsection{Formal properties of \mlts}\label{ssec:formal}

Given the restrictions of Section~\ref{ssec:unification}, we can list
the following three formal properties about \mlts.

\begin{theorem}[Nominals do not escape]
Let $E$ be the abstract syntax of an \mlts program that does not
contain any free nominal.  If\/ $\vdash\eval{E}{V}$ is provable then
$V$ does not contain any free nominals.
\end{theorem}

The proof of this follows from a simple induction on the structure of
proofs in the logic $\Gee$: the precise nature of the semantic
specification given in Figure~\ref{fig:natsem} is not relevant.
The systematic use of the $\nabla$-quantifier guarantees this
conclusion.

\begin{theorem}[Type preservation]
If the typing judgment $\vdash E : A$ and the evaluation
judgment $\vdash\eval{E}{V}$ holds, then so does $\vdash V : A$.
\end{theorem}

The proof is mostly standard, but it must handle the pattern-matching
rule defined by nominal abstraction. This is done using our rigid
paths $\pi$. We can easily prove that if the judgment
$\clause V {\nab {(\lambda Z_1. \dots \nab (\lambda Z_n. \ubranch
    p N))}}$ holds, then the path $\pi_i$ of $Z_i$ in $p$ is also the
path of some nominal $Y_i$ in $v$. Then one needs an intermediate
lemma to say that the type $A$ of a value or pattern and a path $\pi$
within that value or pattern uniquely determine the type $B$ of the
sub-value or the sub-pattern; because $p$ and $v$ have the same type,
the nominals $Z_i$ and $Y_i$ must also have the same type, which is
key to the type-preservation argument.

\begin{theorem}[Determinacy of evaluation]
\label{thm:determinacy}
If\/ $\vdash\eval{E}{V}$ and $\vdash\eval{E}{U}$ then $V = U$.
\end{theorem}

The proof of this theorem follows the usual outline. Again, rigid
paths are used in the pattern-matching rule to justify that the
nominals bound by nabla-abstraction are uniquely determined.

Detailed proofs of these theorems can be found in the forthcoming
Ph.D. dissertation of the first author \cite{gerard19phd}.

\section{Binder mobility}
\label{sec:mobility}


We started this programming language project with the desire to treat
binders in syntax as directly and naturally as possible.
We approached this project by designing the \mlts language with
more binders than, say, OCaml: it has not only
the usual binders for building functions and for refactoring
computation (via the \lsti{let} construction) but also new
binders that are directly linked to binders in data (via the
\lsti{new X in}, \lsti{nab X in}, and \lsti{X\ } operators).
Finally, the natural semantics of \mlts in $\Gee$ and its
implementation in \lP are all based on using logics that contain rich
binding operators that go beyond the usual universal and existential
quantifiers.
It is worth noting that if one were to write \mlts programs that do
not need to manipulate data structures containing bindings, then the
new binding features of \mlts would not be needed and neither would
 the novel
features of both $\Gee$ and \lP.
Thus, in a sense, binders have not been formally implemented in this
story: instead, binders of one kind have been implemented and specified
using binders in another system.
We were able to complete a prototype implementation of \mlts since the
implementers of \lP provide a low-level implementation of bindings
that we are able to use in our static and dynamic semantics
specifications.

One way to view the processing of a binder is that one first \emph{opens} the
abstraction, processes the result (by ``freshening'' the freed
names), and then \emph{closes} the abstraction~\cite{pottier06ml}.
In the setting of \mlts, it is better to view such processing as the
\emph{movement} of a binder: that is, the binder in a data
structure actually gets re-identified with an actual binder in the
programming language.
As we illustrated in Section~\ref{sec:examples} with the following
step-by-step evaluation
\begin{lstlisting}
size (Abs (X\ (Abs (Y\ (App(X,Y))))));;
new X in 1 + size (Abs (Y\ (App(X,Y))));;
new X in 1 + new Y in 1 + size (App(X,Y));;
new X in 1 + new Y in 1 + 1 + size X + size Y;;
new X in 1 + new Y in 1 + 1 + 1 + 1;;
\end{lstlisting}
the bound variable occurrences for \lsti{X} and \lsti{Y} simply move.
It is never the case that a bound variable becomes free:
instead, it just becomes bound elsewhere.
%

\subsection{$\beta_0$ versus $\beta$}
\label{ssec:beta0}

As we describe in Section~\ref{sssec:hop}, we insist that in the
left side of a match rule, all subexpressions of the form \lsti{(r @ X1 ... Xn)}
are such that the scope of the binding for \lsti{r} contains the scopes
of the bindings for the distinct variables in \lsti{X1}, $\ldots$,
\lsti{Xn}.
On the right-hand side of a match rule, however, it seems that one has
an interesting choice.
If on the right, we have an expression of the form
\lsti{(r @ t1 ... tn)} then clearly, the terms \lsti{t1}, $\ldots$, \lsti{tn} are
intended to be substituted into the abstraction that is instantiated for
the pattern variable \lsti{r}: that is, we need to use
$\beta$-conversion on this redex.
One design choice is that we restrict the terms \lsti{t1}, $\ldots$,
\lsti{tn} to be distinct nominals just as on the left-hand-side: in
this case, $\beta$-reduction of the expression \lsti{(r @ t1 ... tn)}
requires only $\beta_0$ reductions.
A second choice is that we allow the terms \lsti{t1}, $\ldots$,
\lsti{tn} to be unrestricted: in this case, $\beta$-reduction of the
expression \lsti{(r @ t1 ... tn)} requires more general (and costly)
$\beta$-reductions.
%
%
%
Our current implementation allows for these richer forms of \lsti{@}
expressions.

A similar trade-off between allowing $\beta$-conversion or just
$\beta_0$ conversion has also been studied within the theory and
design of the $\pi$-calculus.
In particular, the full $\pi$-calculus allows the substitution of
arbitrary names into input prefixes (modeled by $\beta$-conversion)
while the $\pi_I$-calculus ($\pi$-calculus with internal mobility
\cite{sangiorgi96tcs}) is restricted in such a way that the only
instances of $\beta$-conversions are, in fact, $\beta_0$-conversions
(see Chapter 11 in~\cite{miller12proghol}).

Another reason to identify the $\beta_0$ fragment of
$\beta$-conversion is that $\beta_0$ reduction provides support for
binder mobility and it can be given effective implementations,
sometimes involving only constant time operations (see
Section~\ref{ssec:cost}).

\subsection{Nominal-escape checking}
\label{sec:escape}

As we have mentioned in Section~\ref{sec:natsem}, nominals are not
allowed to escape their scope during evaluation and quantifier
alternation can be used to enforce this restriction at the logic
level.
When one implements the logic, one needs to implement (parts of) the
unification of simply typed $\lambda$-terms~\cite{huet75tcs} and such
unification is constantly checking that bound variable scopes are
properly restricted.
There are times, however, when the expensive check for escaping
nominals are not, in fact, needed.
In particular, it is possible to rewrite the inference rule in
Figure~\ref{fig:natsem} for the \lsti{new} binding operator as the
following rule.
\[
\infer{\vdash\eval{\new~E}{V}}
      {\vdash\nabla X.\eval{(E~X)}{(U~x)}\qquad U = \lambda X. V}
\]
Here, both $U$ and $V$ are quantified universally around the inference
rule.
Attempting a proof of the first premise can result in the construction
of some (possibly large) value, say $t$, such that
$\vdash\eval{(E~X)}{t}$ holds.
We can immediately form the binding of $U\mapsto\lambda X.t$ without
checking the structure of $t$.
The second premise is where the examination of $t$ may need to take
place: if $X$ is free in $t$, then there is no substitution for $V$
that makes $\lambda X.t$ equal to $\lambda X. V$.
This check can be expensive, of course, since one might in principle
need to examine the entire structure of $t$ to solve this second
premise.
There are many situations, however, where such an examination is not
needed and they can be revealed by the type system.
For example, if the type of $U$ is, say, \lsti{tm => int}, there
should not be any possible way for an untyped $\lambda$-term to have
an occurrence inside an integer.
Furthermore, there are static methods for examining type declarations in order to
describe if a type $\tau_1\ra\tau_2$ (for primitive types $\tau_1$ and
$\tau_2$) can be inhabited by at most vacuous $\lambda$-terms (see, for
example,~\cite{miller92jsc}).
Of course, if the types of $\tau_1$ and $\tau_2$ are the same (say,
\lsti{tm}), then type information is not useful here and a check of
the entire structure $t$ might be necessary.
Other static checks and program analyses might be possible as a way to
reduce the costs of checking for escaping nominals: the paper
\cite{pottier07lics} includes such static checks albeit for a
technically different functional programming language, namely
FreshML~\cite{gabbay03icfp}.


\subsection{Costs of moving binders}\label{ssec:cost}


As we have mentioned before, binders are able to move from, say, a
term-level binding to a program-level binding by the use of $\beta_0$.
In particular, if $y$ is a binder that does not appear free in the
abstraction $\lambda x.B$ then the $\beta_0$ reduction of $(\lambda
x.B)y$ causes the $x$ binding in $B$ to move and to be identified with the
$y$ binder in $B[y/x]$.
If one must actually do the substitution of $y$ for $x$ in $B$, a
possibly large term (at least its spine) must be copied.
However, there are some situations where this movement of a binding
can be inexpensive.
For example, consider again the following match rule for \lsti{size}.
\begin{lstlisting}
   | Abs(r) -> 1 + (new X in size (r @ X))
\end{lstlisting}
If we assume that the underlying implementation of terms use De
Bruijn's indexes, it is possible to understand the rewriting
needed in applying this match clause to be a constant time operation.
In particular, if \lsti{r} is instantiated with an abstraction then
its top-level constructor would indicate where a binder of value $0$
points.
If we were to compile the syntax \lsti{(r @ X)} as simply meaning that
that top-level constant is stripped away, then a binder of value $0$
in the resulting term would automatically point (move) to being bound
by the \lsti{new X} binder.
While such a treatment of binder mobility without doing substitution is
possible in many of our examples, it does not cover all cases.
In general, a more involved scheme for implementing
binder mobility must be considered.
This kind of analysis and implementation of binder mobility is used in
the ELPI implementation of $\lambda$Prolog~\cite{dunchev15lpar}.

\section{Future work}
\label{sec:future}

There is clearly much more work to do.
While the examples presented in this paper illustrate that the new
features in \mlts can provide elegant and direct support for computing
with binding structures, we plan to develop many more examples
centered on the general area of implementing theorem provers and
compiler construction.
A more effective implementation is also something we wish to target
soon.
It seems likely that we will need to consider extensions to the usual
abstract machine models for functional programming in order to get
such a direct implementation.
A first step in this direction would be to first design a small-step
(SOS) semantics equivalent of our natural semantics.

The cost of basic operations in \mlts must also be understood better.
As we noted in Section~\ref{sec:examples}, we could design pattern
matching in clauses in such a way that they might require the
recursive descent of entire terms in order to know if a match was
successful.
The language could also be designed so that such a costly check is
never performed during pattern matching: for example, one could insist
that every pattern variable is \lsti{@}-applied to a list of
\emph{all} nominal abstractions that are in the scope of the binding
for that pattern variable.
In that case, a recursive descent of terms is not needed.

Given the additional expressivity of \mlts, the usual static checks
used to produce warnings for non-exhaustive matchings are missing
cases that we should add.
As mentioned in Section~\ref{sec:mobility}, still other static checks are
needed to help a future compiler avoid making costly checks.
%
Finally, adding polymorphic typing should be possible following the
pattern already established by OCaml.

It is also interesting to see to what extent binders interact with a
range of non-functional features, such as references.
A natural starting point to explore the possible interaction of
effectful features would be to use a natural semantics treatment based
on linear logic (see, for example,~\cite{chirimar95phd,miller96tcs}):
the logical features of $\Gee$ should also work well in a linear logic
setting.

Finally, the treatment of syntax with bindings generally leads to the
need to manipulate contexts and association lists that relate bindings
to other bindings, to types, or to bits of code.
We have already seen association lists used in
Figure~\ref{fig:simple}. 
It seems likely that more sophisticated \mlts examples will require
singling out contexts for special treatment.
Although the current design of \mlts does not commit to any special
treatment of context, we are interested to see what kind of treatment
will actually prove useful in a range of applications.

\section{Related work}
\label{sec:related}

The term \emph{higher-order abstract syntax (HOAS)} was
introduced in~\cite{pfenning88pldi} to describe an encoding technique
available in \lP.
A subsequent paper identified HOAS as a technique ``whereby variables
of an object language are mapped to variables in the
metalanguage''~\cite{pfenning99cade}.
When applied to functional programming, this description
implies the mapping of bindings in syntax to the bindings that
create functions.
Unfortunately, such encoding technique often lacks adequacy (since
``exotic terms'' can appear~\cite{despeyroux95tlca}), and structural
recursion can slip away~\cite{gabbay99lics}.
The terms \emph{\lts}~\cite{miller99surveys,miller18jar} and
\emph{binder mobility}~\cite{miller04csl} were later introduced to describe the
different and more syntactic approach that we have used here.

\subsection{Systems with two arrow type constructors}

The $ML_\lambda$~\cite{miller90tr} extension to ML is similar to \mlts
in that it also contains two different arrow type constructors
(\lsti{->} and \lsti{=>}) and pattern matching was extended to allow
for pattern variables to be applied to a list of distinct bound
variables.
The \lsti{new} operator of \mlts could be emulated by using the
backslash operator and the ``discharge'' function of $ML_\lambda$.
Critically missing from that language was anything similar to the
\lsti{nab} binding of \mlts.
Also, no formal specification and no implementation were ever offered.
Licata \& Harper~\cite{licata09icfp} have used the universe feature of
Agda 2 to provide an implementation of bindings in data structures
that also relies on supporting two different implications-as-types.

Nominals and nominal abstraction, in the sense used in this paper, were
first conceived, studied, and implemented as part of the Abella theorem
prover~\cite{baelde14jfr}.
%
%
While Abella only has one arrow type constructor, that arrow type
maps to the \lsti{=>} of \mlts: this is possible in Abella
since computation is performed at the level of relations and not
functions.
As a result, the function type arrow \lsti{->} of \mlts and OCaml is
not needed.
Thus the distinction mentioned in~\cite{licata09icfp} between an arrow
for computation and an arrow for binding is, in fact, also present in
Abella, although computations are not represented functionally.

\subsection{Systems with one arrow type constructor}


The Delphin design is probably the closest to
\mlts, in particular~\cite{schurmann05tlca} introduced
a programming-language version of the $\nabla$ quantifier from
\cite{miller05tocl}, whose usage is related to the $\nabla$ of
\mlts. In Delphin, $\nabla$ introduces normal term variables (there is
no separate class of nominal constants), while $\mlts$ presents
nominals as closer to datatype constructors, with a natural usage in
pattern-matching.

Delphin makes nominal-escape errors impossible at runtime by imposing
a static discipline to prevent them, while $\mlts$ allows runtime
failure in order to allow for more experimentation. The original
proposal in~\cite{schurmann05tlca} uses a type modality that imposes
a strict FIFO discipline on free variables. This discipline was found
too constraining;~\cite{poswolsky08esop} completely eschews
a \lsti{new} construct (its $\nu x.\,e$ binder actually corresponds to
nominal abstraction \lsti{X\e} in \mlts), and~\cite{poswolsky08lfmtp}
uses a type-based restriction (type subordination), only allowing to
introduce a fresh nominal in expressions whose return types only
contains values that cannot contain this nominal. This discipline
accepts some examples from our paper, for example \lsti{size} in
Figure~\ref{fig:size} and \lsti{id} in Figure~\ref{fig:simple}, but
rejects other (safe) programs, such as the second and third one-liner
examples of Section~\ref{sec:hop}.  Richer static disciplines have been
proposed for FreshML~\cite{pitts00mpc,pottier07lics},
but they add complexity, and interact poorly with the
introduction of mutable state; \mlts is an experimental design aiming
for expressivity, so we decided to allow dynamic escape failures
instead.

Beluga~\cite{pientka10ijcar} allows the programmer to use both
dependent types and recursive definitions as well as an integrated
notion of context (along with a method to describe certain invariants
using \emph{context schema}).
Static checks of Beluga programs can be used to prove the formal
correctness of Beluga programs (commonly by proving that a given
piece of program code is, in fact, a total function).
As a result, a checked Beluga program is often a formal proof.
Since a wide range of formal systems can be encoded naturally using
dependently typed $\lambda$-terms~\cite{avron92jar,harper93jacm},
Beluga programs can be used for both programming with and reasoning
about the meta-theory of those formal systems.
Since bindings and contexts are part of the vocabulary of Beluga,
these formal proofs can capture the metatheory of logical and
computational systems (such as natural deduction proof systems and
the operational semantics of rich programming languages).
The goal of \mlts is intended only to support programming and not
directly reasoning: the intent of the new features of \mlts is only to
support the manipulation of syntax containing bindings.
A possibly interesting comparison between \mlts and Beluga might be
explored by using typing and contexts in the latter in a mostly
trivial way.
It is likely that Beluga could code most \mlts programs although using
different primitives.

\subsection{Systems using nominal logic}

The FreshML~\cite{gabbay03icfp} and C$\alpha$ML~\cite{pottier06ml}
functional programming languages provide an approach to names based on
nominal logic~\cite{pitts03ic}.
These two programming languages provide for an abstract
treatment of names and naming.
Once naming is available, binding structures can also be implemented.
In a sense, the design of these two ML-variants are also more
ambitious than the design goal intended for \mlts: in the latter, we
were not focused on naming but just bindings.

The recent paper~\cite{ferreira17esop} introduces a syntactic
framework that treats bindings as primitives.
That framework is then integrated with various tools and with the
framework of contextual types (similar to that found in Beluga)
in order to provide a programmer of, say, OCaml with sophisticated
tools for the manipulation of syntax and binders.
A possible future target for \mlts could be to provide
such tools more directly in the language itself.

\subsection{Challenge problems and benchmarks}

Genuine comparisons between different programming languages are
generally hard to achieve.
For example, in the area of logical frameworks and related theorem
provers, there are also a number of formal systems and computer
implementations.
In order to understand the relative merits of these different systems,
challenge problems and benchmarks~\cite{aydemir05tphols,felty15jar}
have been proposed to help people sort out specific merits and
challenges of one system relative to another.
In depth comparisons of the programming languages described above will
probably require similar in-depth comparisons on representative
programming tasks.

\section{Conclusion}
\label{sec:conclusion}

While the \lts approach to computing with syntax containing bindings
has been successfully developed within the logic programming setting
(in particular, in \lP and Twelf), we provide in this paper another
example of how binding can be captured in a functional programming
language.
Most of the expressiveness of \mlts arises from its increased use of
program-level binding.
The sophistication needed to correctly exploit binders and quantifiers
in \mlts is a skill most people have learned from using quantification
in, for example, predicate logic.

We have presented a number of \mlts programs and we note that
they are both natural and unencumbered by concerns about managing
bound variable names.
We have also presented a typing discipline for \mlts as well as a
formal specification of its natural semantics: this latter task was
aided by being able to directly exploit a rich logic, called $\Gee$,
that allows capturing both \lts and binder mobility.
Finally, the natural semantics specification and the typing system
were directly implementable in \lP.
A prototype implementation is available for helping to judge the
expressiveness of \mlts programs.

\smallskip
\noindent{\bf Acknowledgments.}  We thank Kaustuv Chaudhuri, Fran\c
cois Pottier, Enrico Tassi, the HOPE Workshop 2018 audience, and the
anonymous reviewers for their helpful comments and observations.

\pagebreak


\end{document}